\documentclass[fleqn,usenatbib,useAMS]{mnras}

\usepackage{graphicx}	% Including figure files
\usepackage{amsmath}	% Advanced maths commands
\usepackage{amssymb}	% Extra maths symbols
\usepackage{multicol}        % Multi-column entries in tables
\usepackage{bm}		% Bold maths symbols, including upright Greek
\usepackage{pdflscape}	% Landscape pages

\usepackage[T1]{fontenc}
\usepackage{ae,aecompl}

\usepackage{newtxtext,newtxmath}
\title[Position change of four radio sources]{Unprecedented change in the position of four radio sources}

\author[Oleg Titov et al.]{
Oleg Titov,$^{1}$
S\'{a}ndor Frey,$^{2,3}$
Alexey Melnikov,$^{4}$
S\'{e}bastien Lambert,$^{5}$
Fengchun Shu,$^{6}$
Bo Xia,$^{6}$
\newauthor
Javier Gonz\'{a}lez,$^{7}$
Bel\'{e}n Tercero,$^{7}$
Sergey Gulayev,$^{8}$
Stuart Weston,$^{8}$
and Tim Natusch$^{8}$
\\
$^{1}$Geoscience Australia, PO Box 378, Canberra, ACT 2601, Australia\\
$^{2}$Konkoly Observatory, ELKH Research Centre for Astronomy and Earth Sciences, Konkoly Thege Mikl\'os \'ut 15-17, H-1121 Budapest, Hungary\\
$^{3}$Institute of Physics, ELTE E\"otv\"os Lor\'and University, P\'azm\'any P\'eter s\'et\'any 1/A, H-1117 Budapest, Hungary\\
$^{4}$Institute of Applied Astronomy, Russian Academy of Sciences, Kutuzova Embankment 10, St. Petersburg, 191187, Russia\\
$^{5}$SYRTE, Observatoire de Paris -- Université PSL, CNRS, Sorbonne Université, LNE, 61 avenue de l'Observatoire, 75014, Paris , France\\
$^{6}$Shanghai Astronomical Observatory, Chinese Academy of Sciences, 80 Nandan Road, Shanghai 200030, China\\
$^{7}$Observatorio de Yebes (IGN), Apartado 148, 19180 Yebes, Spain\\
$^{8}$Institute for Radio Astronomy and Space Research, 
Auckland University of Technology, 120 Mayoral Drive, 
Auckland 1010, New Zealand}

\pubyear{2021}

\begin{document}
\label{firstpage}
\pagerange{\pageref{firstpage}--\pageref{lastpage}}
\maketitle

\begin{abstract}
Astrometric positions of radio-emitting active galactic nuclei (AGNs) can be determined with sub-milliarcsec accuracy using very long baseline interferometry (VLBI). The usually small apparent proper motion of distant extragalactic targets allow us to realize the fundamental celestial reference frame with VLBI observations. However, long-term astrometric monitoring may reveal extreme changes in some AGN positions. Using new VLBI observations in 2018--2021, we show here that four extragalactic radio sources (3C\,48, CTA\,21, 1144+352, 1328+254) have a dramatic shift in their positions by 20--130 milliarcsec over two decades. For all four sources, the apparent positional shift is caused by their radio structure change.
\end{abstract}


%%%%%%%%%%%%%%%%%%%%%%%%%%%%%%%%%%%%%%%%%%%%%%%%%%

%%%%%%%%%%%%%%%%% BODY OF PAPER %%%%%%%%%%%%%%%%%%

% The MNRAS class isn't designed to include a table of contents, but for this document one is useful.
% I therefore have to do some kludging to make it work without masses of blank space.
%\begingroup
%\let\clearpage\relax
%\tableofcontents
%\endgroup
%\newpage

\section{Introduction}
\label{sec:intro}

The International Celestial Reference System (ICRS) is the standard quasi-inertial reference system with origin at the Solar System Barycenter (SSB) and with the fundamental axes which are “fixed” with respect to the selected fiducial points on the sky. The International Celestial Reference Frame (ICRF) realizes the ICRS by the positions of 4536 extragalactic radio sources that are assumed to have no measurable proper motion.  The recent ICRF realization, known as ICRF3 \citep{2020A&A...644A.159C}, reaches an accuracy of ~30 microarcsec ($\mu$as) in both coordinate directions (right ascension, RA, and declination, Dec). The fundamental axes are fixed by positions of 303 so-called ICRF3 “defining” radio sources. Astrometric very long baseline interferometry (VLBI) observations regularly performed since the late 1970s \citep{1998RvMP...70.1393S} are based on group delay measurements on long, intercontinental baselines using bandwidth synthesis at standard frequencies around 2.3 and 8.4 GHz ($S$ and $X$ bands). The structure of most radio active galactic nuclei (AGNs) is dominated by compact emission on milliarcsec (mas) or sub-mas angular scales, making them suitable astrometric reference objects. The location of these celestial reference points is affected by the actual brightness distribution, but for nearly unresolved compact sources it is basically determined by the brightness peak at $X$ band \citep{2000ApJS..128...17F}. New VLBI observing programs are currently coordinated by the International VLBI Service for Geodesy and Astrometry \citep[IVS,][]{2017JGeod..91..711N}.

The \textit{Gaia} space mission  \citep{2021A&A...649A...1G}  produced an independent catalogue of the AGNs observable in optical wavelengths, with substantial overlap between the \textit{Gaia} and ICRF3 catalogues. Some authors have found that the optical and radio positions are not perfectly aligned for many of the common objects \citep[e.g.][]{2016A&A...595A...5M, Makarov2017,2019ApJ...873..132M,2019MNRAS.482.3023P}, therefore, further work is to be done to improve the consistency between the radio and optical realizations of the ICRS. Radio reference frame source coordinates refer to the positions of the compact radio-emitting features in the core and/or jet. The optical counterparts are often associated with extended galaxies that may cause their optical centroids to not be aligned with the radio-emitting features. In some cases, the optical counterparts are not even detected with the existing ground-based facilities due to faint emission in optical wavelengths \citep{2014AJ....147...95Z}. 

Some of the radio reference frame sources demonstrate a moderate astrometric instability within $1-2$~mas \citep[e.g.][]{2003A&A...403..105F, 2007AstL...33..481T}. More fundamental astrophysical studies have found that the frequency-dependent core shift observed with VLBI in parsec-scale jets may lead to an astrometric effect as much as 1.4~mas \citep{2008A&A...483..759K}, and an average shift between the $X$-band radio and optical astrometric positions is expected to be approximately 0.1~mas. Sometimes, radio reference frame sources display a persistent evolution in coordinates with time that could be approximated by a linear function and interpreted as an apparent proper motion. This observed proper motion usually agrees well with the superluminal motion of the brightest component detected in VLBI images and may reach 1~mas\,yr$^{-1}$ over a short period of time \citep[e.g.][]{ 1993ApJ...402..160A, 1997AJ....114.2284F, 2003A&A...403..105F, 2004ApJ...609..539K, 2007AstL...33..481T, 2011A&A...529A..91T, 2011AJ....141..178M}. These astrometrically unstable objects are monitored with geodetic VLBI observations, and the list of the radio sources unsuitable for establishing of fundamental ICRS axes is updated in each ICRF realisation \citep{Ma_1998,Fey_2015,2020A&A...644A.159C}. In the most recent revision, known as ICRF3, based on VLBI observations from 1979 to spring 2018, \citet{2020A&A...644A.159C} did not detect an extremely large offset in radio source positions.

An issue, however, is that since the recent ICRF3 system was officially adopted by the International Astronomical Union (IAU) XXII General Assembly in 2018, several extragalactic radio sources have been discovered, whose apparent position has changed by $20-130$~mas or even larger on a timescale of $3-20$~yr \citep[e.g.][]{2020RNAAS...4..108T,2021RNAAS...5...60F}. Such unforeseen dramatic changes of the position cause serious problems for VLBI data analysts because none of the standard geodetic software packages are designed to handle them. A substantial loss in the accuracy of the geodetic parameters (Earth orientation parameters, geocentric positions of the VLBI telescopes, etc.) may happen, if these dramatic astrometric perturbations are ignored.

Finally, the apparent proper motions of the reference radio sources may contain some tiny systematic effects. For example, the Sun's Galactocentric acceleration induces the effect of secular aberration drift that is observed as the dipole systematic of $5-7 ~\mu$as\,yr$^{-1}$ \citep{2011A&A...529A..91T, 2013A&A...559A..95T, 2019A&A...630A..93M, 2020A&A...644A.159C}. Moreover, the  primordial gravitational waves could be also detected, hypothetically, as a quadrupole systematic pattern around the sky \citep[e.g.][]{1966ApJ...143..379K, 1997ApJ...485...87G, 2011A&A...529A..91T}. Individual proper motions of all radio sources are used to estimate this tiny systematic effect. Any large positional offset, if being unaccounted for, would destroy pattern and bias the estimates of the systematic effects. Therefore, all the radio sources with large displacement and unreasonably high positional displacement rates need to be identified and deselected a priori.

%\begin{verbatim}
%\documentclass{mnras}
%\end{verbatim}
%Then compile \LaTeX\ (and if necessary \bibtex) in the usual way.

\begin{table}
 \caption{The list of the VLBI sessions of the RUA project.}
 \label{table_observations}
 \begin{tabular}{cccc}
  \hline
  Session & Date & Radio & Observed \\ [1ex]
code & YYYY-MM-DD  & telescopes & targets \\  [1ex]
  \hline
  \hline
RUA038 & 2020-06-27 & Bd-Sh-Sv-Ys-Zc & 1144+352 \\ [1ex]
RUA039 & 2020-11-28 & Bd-Sv-Ys-Zc & 1328+254 \\ [1ex]
RUA040 & 2020-12-26 & Bd-Sh-Sv-Ys-Zc & 1328+254  \\ [1ex]
RUA041 & 2021-03-20 & Bd-Sh-Sv-Zc & CTA\,21 \\ [1ex]
RUA042 & 2021-03-27 & Bd-Sh-Sv-Wa-Zc & CTA\,21 \\ [1ex]
RUA043 & 2021-04-10 & Bd-Sh-Sv-Zc & CTA\,21, 1328+254 \\ [1ex]
  \hline
 \end{tabular}
\end{table}

\section{Observations and data reduction}
\label{sec:obs}

In this section, we describe the VLBI observations and data reduction pipeline used in this paper.

\begin{table}
\caption{List of four radio sources which demonstrate a large positional displacement. The flux densities are given at 8.4~GHz.}
\vspace{3mm}
\begin{tabular}{l c c c }
 \hline
 Source & Approximate flux & Redshift   & Period of observations \\ [1ex]
 &  density (mJy) &  & (years)  \\ [1ex]
  \hline \hline
  3C\,48   & 100 & 0.367 & 1990--2019   \\  [1ex] 
  CTA\,21  & 100 & 0.907 & 1996--2021   \\ [1ex] 
  1144+352 &  50 & 0.063 & 1996--2020   \\  [1ex] 
  1328+254 &  70 & 1.055 & 2006--2021   \\  [1ex]
 \hline
\label{table_sources}
\end{tabular}
\end{table}

%\begin{landscape}
\begin{table*}
\caption{Astrometric positions of the four radio sources.}
%\vspace{3mm}
\label{table_source_VLBI_positions}
\begin{tabular}{lcccc}
 \hline
 Source & ICRF3 & Aus2020a.crf & Post-ICRF3 & Number of ``new'' \\ 
 & RA ($^\mathrm{h}$ $^\mathrm{min}$ $^\mathrm{s}$), Dec ($\degr$ $\arcmin$ $\arcsec$) & RA ($^\mathrm{h}$ $^\mathrm{min}$ $^\mathrm{s}$), Dec ($\degr$ $\arcmin$ $\arcsec$) & coordinates & experiments \\ [1ex]
 \hline \hline
  0134+329      & 01 37 41.299545    & 01 37 41.299597    & 01 37 41.299702  & 20 \\  
  (3C\,48) & 33 09 35.13417 & 33 09 35.08787 &  33 09 35.07804 & \\  \hline
  0316+162   & 03 18 57.803043    & 03 18 57.802822    & 03 18 57.803187  & 7 \\  
  (CTA\,21) & 16 28 32.67823 & 16 28 32.68462 &  16 28 32.66443 & \\  \hline
 1144+352  & 11 47 22.129282 & 11 47 22.129307 & 11 47 22.129038 & 2 \\   
  & 35 01 07.53082 & 35 01 07.53354 &  35 01 07.53494 & \\      \hline
  1328+254           & 13 30 37.695195     & 13 30 37.695979    & 13 30 37.695997  & 4 \\ 
  (3C\,287) & 25 09 10.94444  & 25 09 10.97397  & 25 09 10.98847 & \\
  \hline
  \end{tabular}
\\
Notes: Col.~2 -- official positions in the ICRF3 catalogue; Col.~3 -- intermediate positions in the aus2020a.crf solution used to obtain the “mean” coordinates shown in Col.~4 which are based on the VLBI data that were not applied for the ICRF3 catalogue solution.
\end{table*}

\begin{table*}
\caption{Optical astrometric positions of the three of four radio sources have optical counterparts observed: Gaia DR1 (second column), Gaia EDR3 (third column), and the difference between them (fourth column), $\Delta\alpha$ and $\Delta\delta$.}
\label{table_source_optical_VLBI_positions}

\begin{tabular}{l c c c }
 \hline
 Source & Gaia DR1 & Gaia EDR3 & Gaia EDR3 $-$ Gaia DR1   \\ 
 & RA ($^\mathrm{h}$ $^\mathrm{min}$ $^\mathrm{s}$), Dec ($\degr$ $\arcmin$ $\arcsec$) & 
 RA ($^\mathrm{h}$ $^\mathrm{min}$ $^\mathrm{s}$), Dec ($\degr$ $\arcmin$ $\arcsec$) & (mas) \\ [1ex]
 \hline \hline
  0134+329   & 01 37 41.299595   & 01 37 41.299585    & $ -0.15 $ \\  
  (3C48) & 33 09 35.07995 & 33 09 35.07913 & $ -0.08 $ \\ \hline
  1144+352  & 11 47 22.129358 & 11 47 22.129212 &  $-2.19$ \\
  & 33 01 07.53309 & 35 01 07.53475 &  $+1.66$  \\ \hline
  1328+254 & 13 30 37.695773  & 13 30 37.695738  &  $-0.53$\\
  (3C\,287) & 25 09 10.98405 & 25 09 10.98424  &  $+0.19$ \\
  \hline

\end{tabular}
\end{table*}

\begin{table*}
\caption{Radio--radio and radio--optical offsets of the four radio sources.}
\label{table_radio_optical}
%\vspace{3mm}
\begin{tabular}{lccc}
 \hline
 Source & Post-ICRF3 $-$ ICRF3 & \textit{Gaia} DR1 $-$ ICRF2 & \textit{Gaia} EDR3 $-$ ICRF3   \\ 
 &  (mas) & (mas) & (mas)\\ [1ex]
 \hline \hline
  0134+329  &  0.16   & 1.44 & 0.62 \\  
  (3C48) &  $ -56.13 $ & $-53.80 $ & $-55.02 $ \\ \hline
  0316+162  & 0.14    & ... & ... \\  
 (CTA\,21)  & $-13.80$ & ... & ... \\  \hline
    1144+352   & $-0.24$ & $-18.2$ & $-1.05$ \\
   &  4.12 &  10.59 &  4.02 \\ \hline
  1328+254    & 0.85   & 80.96 & 8.15\\
  (3C\,287)   & 44.02  &  105.49 & 39.80 \\
  \hline
\end{tabular}
\\
Notes: Col.~2 -- the difference $\Delta\alpha$ and $\Delta\delta$ between the ICRF3 official positions and the new positions given in the 4th column of Table~\ref{table_source_VLBI_positions} for RA and Dec, separately; Col.~3 -- the difference between the \textit{Gaia} DR1 positions and the ICRF2 positions published; Col.~4 --  the total difference between the \textit{Gaia} EDR3 positions and the ICRF3 positions.
\end{table*}

All observing sessions before 2020 were part of the permanent VLBI program coordinated by the IVS \citep{2017JGeod..91..711N}. Most experiments in the 1990s were performed with astrometric networks including radio telescopes at such geodetic facilities as Gilmore Creek, Westford, Kauai, NRAO85, NRAO20, Fortaleza, Wettzell, Kokee Park, Algonquin Park, as well as the radio telescopes belonging to the Very Large Baseline Array (VLBA). Since 2000, most of the IVS experiments were done using the VLBA network, with some occasional observations at the geodetic stations. More details are given in the notes on individual radio sources in Sect.~\ref{sec:notes}.

Six non-IVS 24-hour astrometric sessions were run with an ad-hoc global VLBI array involving the Russian National Quasar VLBI Network \citep{SHUYGINA2019150}, the 40-m Yebes radio telescope (Ys) of the Yebes Observatory in Spain, and the 25-m Sheshan radio telescope (Sh) of the Shanghai Astronomical Observatory in China. The  Quasar VLBI Network includes three 32-m radio telescopes Badary (Bd), Svetloe (Sv), and Zelenchukskaya (Zc). One session (27 Mar 2021 (RUA042)) was carried out with involving the 30-m Warkworth radio telescope (Wa) of the Institute for Radio Astronomy and Space Research in New Zealand.
The session names, observation date, participating radio telescopes, and the target sources are presented in  Table~\ref{table_observations}.
%The session names, observation date, duration, frequency bands, participating array, and target source's scan duration on those sessions are presented in the Table~\ref{table_observations}.

All RUA sessions were managed by of the Institute of Applied Astronomy of the Russian Academy of Sciences in Saint-Petersburg. Those VLBI sessions were scheduled using the \textsc{SKED} software \citep{2010ivs..conf...77G} in an astrometric/geodetic fashion, where the number of short scans of the radio sources are distributed over a long time range and the scans of the target sources are mixed with those of the calibrator sources.

Sessions from RUA038 to RUA041 used standard IVS $S/X$-band dual wide-band setup in right circular polarisation. It includes 8 upper side band and 2 lower side band frequency channels (IFs), each with 16~MHz bandwidth in $X$ band, spanning 720~MHz, and 6 upper side band IFs in $S$ band, spanning 140~MHz. The total sampling rate of 1024~Mbit\,s$^{-1}$ was achieved using 2-bit sampling.

Experimental sessions RUA042 and RUA043 were carried out in $X$ band only in right circular polarisation. The setup included 8 upper side band and 8 lower side band IFs, spanning 720 MHz. IF bandwidth of 16 and 32~MHz was used for RUA042 and RUA043, respectively, and the total sampling rate of 1024~Mbit\,s$^{-1}$ and 2048~Mbit\,s$^{-1}$ was achieved using 2-bit sampling, respectively.

The data of the RUA sessions were transferred via the Internet to the Correlator Processing Centre of the Russian Academy of Sciences located in the Institute of Applied Astronomy of the Russian Academy of Sciences in Saint-Petersburg and correlated with the DiFX correlator \citep{2011PASP..123..275D} with 0.5~s integration time and 125~kHz spectral resolution. Post-processing was performed using the \textsc{PIMA} software \citep{2011AJ....142...35P} following the Data analysis pipeline from the \textsc{PIMA} User Guide\footnote{\url{http://astrogeo.org/pima/pima_user_guide.html}}.
Precisely calculated total delays of every observed source in the session were used to estimate the daily coordinates for each 24-hour experiment.

 Radio source coordinates were estimated once per session, together with Earth orientation parameters and station coordinates by the \textsc{OCCAM} 6.3 software \citep{2004ivsg.conf..267T}. The standard procedure to calibrate geodetic VLBI data in accordance with the IERS Conventions 2010 \citep{iers10} was applied. Zenith wet delays with gradients and clock offset instabilities were estimated by the least squares collocation method \citep{2000ITN....28...33T} at each observational moment. The elevation cut-off angle was set to 5\degr. A priori zenith delays were determined from local atmospheric pressure values \citep{1972GMS....15..247S}, which were then mapped to the elevation of the observated sources using the Vienna mapping function \citep[VMF1,][]{2006JGRB..111.2406B}.

\section{Observational results}
\label{sec:res}

We report four radio sources that demonstrate an unprecedented evolution of their positions based on recent astrometric VLBI data obtained since 2018 (Table~\ref{table_sources}). All the sources were included in the list of ICRF3 radio sources \citep{2020A&A...644A.159C} with the coordinates as listed in the second column of Table~\ref{table_source_VLBI_positions}. However, their ICRF3 positions are not consistent with the new observations after 2018, and should be updated.

The third column in Table~\ref{table_source_VLBI_positions} presents the coordinates of the four sources from the aus2020a.crf solution\footnote{\url{https://cddis.nasa.gov/archive/vlbi/ivsproducts/crf/}}. This is one of the recent regular solutions from Geoscience Australia IVS Analysis Centre submitted to the IVS Data Centre in which the coordinates of the four radio sources were estimated as global parameters. It was used as a reference solution to estimate new positions of the four radio sources from each 24-hour geodetic VLBI experiment.

The fourth column in Table~\ref{table_source_VLBI_positions} presents the mean coordinates of each radio source obtained only with the most recent observations not yet used for constructing the ICRF3 catalogue, i.e. after 2017. However, the ICRF3 solution partially comprises the observations of the four sources between 2007 and 2017, i.e. after the substantial evolution since the observations at the first epochs. As a result, the difference between the ICRF3 positions and the positions obtained with new observations (column 2 of Table~\ref{table_radio_optical}) could be smaller than the maximum displacement between the available VLBI observing epochs.

Three of the four radio sources have optical counterparts observed by the \textit{Gaia} mission \citep{2016A&A...595A...1G}, solutions DR1 \citep{2016A&A...595A...2G} and EDR3 \citep{2021A&A...649A...1G}, with coordinates in Table~\ref{table_source_optical_VLBI_positions}. The fourth source, CTA\,21, is below the \textit{Gaia} sensitivity limit.  All three detected sources were reported among the 118 ``dislodged'' AGNs by \citet{Makarov2017}, as they show a significant offset between the ICRF2 positions and their optical \textit{Gaia} DR1 optical counterparts \citep{2016A&A...595A...2G}. The radio--optical offsets between \textit{Gaia} DR1 and ICRF2 are given in the third column of Table~\ref{table_radio_optical} are consistent with the results by \citet{Makarov2017}. It is remarkable that these offsets for all three sources are consistent with the positional change in Table ~\ref{table_comparison}. This may indicate that the radio--optical offset in \citet{Makarov2017} is based on the VLBI observations made before the break only, therefore, the change in positions in Table~\ref{table_comparison} is mostly determined by the positional instability in the radio frequency range, whereas the optical positions are likely to be stable. Indirectly, this conclusion is backed by the tiny difference between the two \textit{Gaia} solutions in the last column of Table~\ref{table_source_optical_VLBI_positions}.

The fourth column of Table~\ref{table_radio_optical} shows the newest radio--optical offsets \textit{Gaia} EDR3 -- ICRF3. They are consistent with the recent results presented by \citet{2021A&A...647A.189X}. As the ICRF3 catalogue comprises some observations of two sources (1144+352 and 1328+254) after the apparent break in positions, their new radio--optical offsets are smaller than the corresponding \textit{Gaia} DR1 -- ICRF2 offsets. Finally, we should note that the previously reported large offsets for 3C\,48 and 1328+254 would drop dramatically, if the correction from the second column of Table~\ref{table_radio_optical} is applied.

\section{Notes on individual radio sources}
\label{sec:notes}

\begin{figure}
\centering
 \includegraphics[width=0.4\textwidth, angle=0]{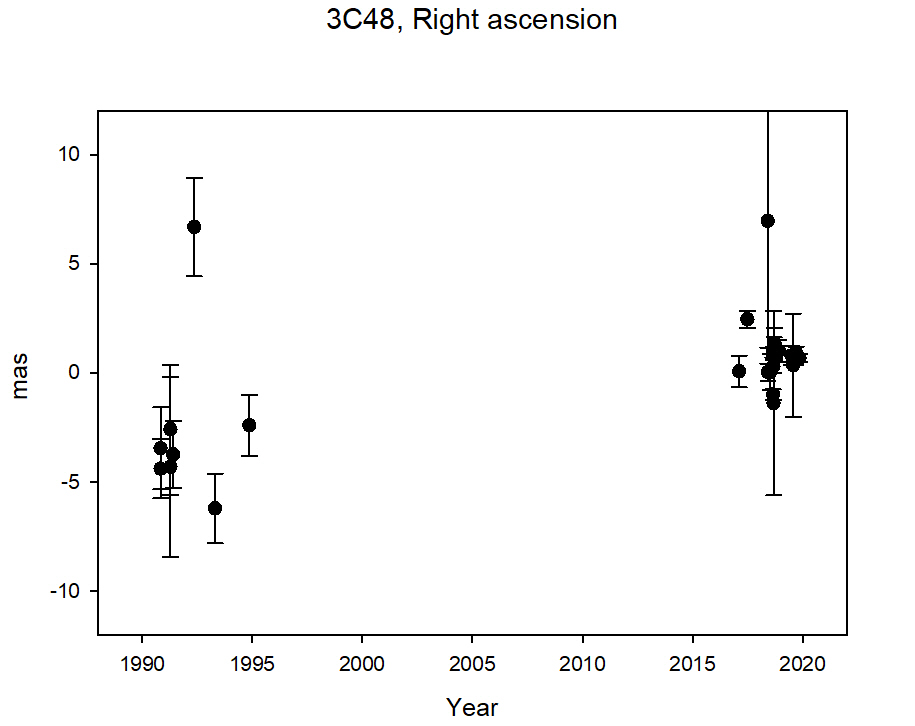}
 \includegraphics[width=0.4\textwidth, angle=0]{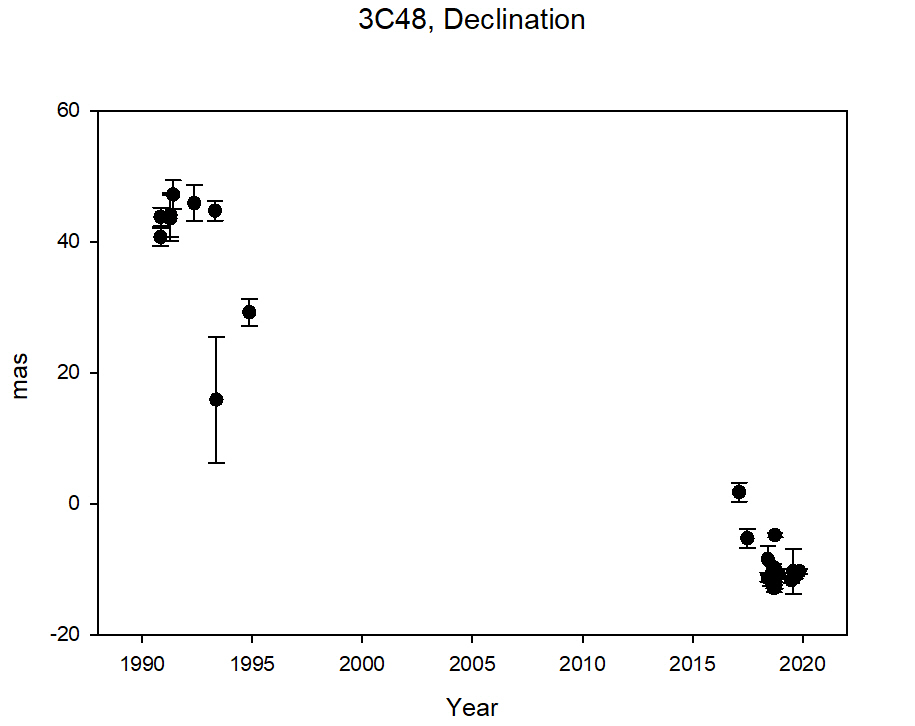}
 \caption{Relative astrometric coordinates of 3C\,48 as a function of time, between 1990 and 2019. The (0,0) point corresponds to the solution aus2020a.crf (Table~\ref{table_source_VLBI_positions}).} 
 \label{fig1}
\end{figure}

\begin{figure}
\centering
 \includegraphics[width=0.4\textwidth, angle=0]{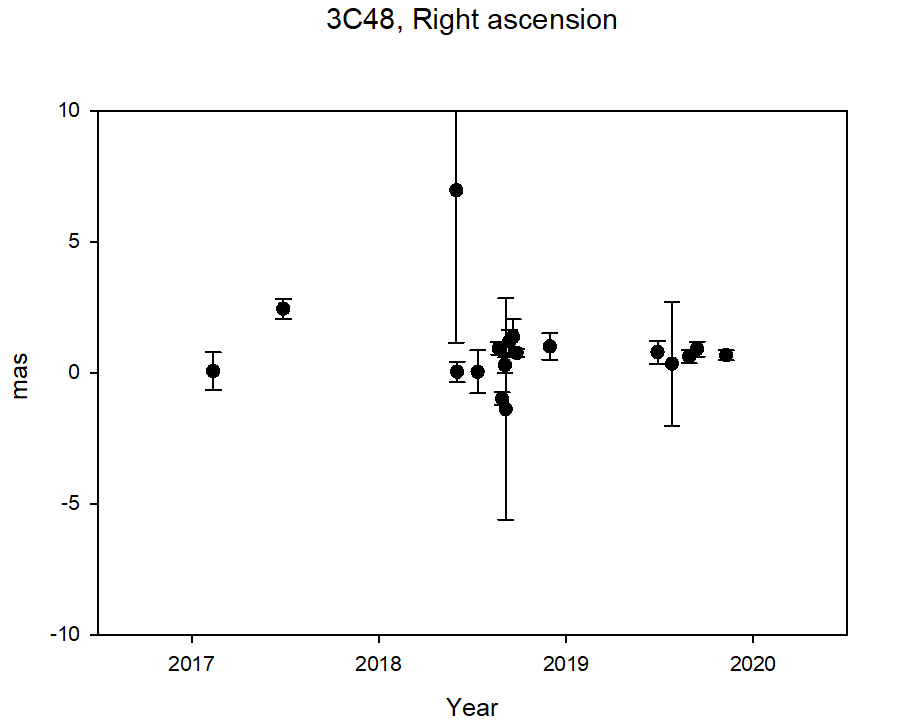}
 \includegraphics[width=0.4\textwidth, angle=0]{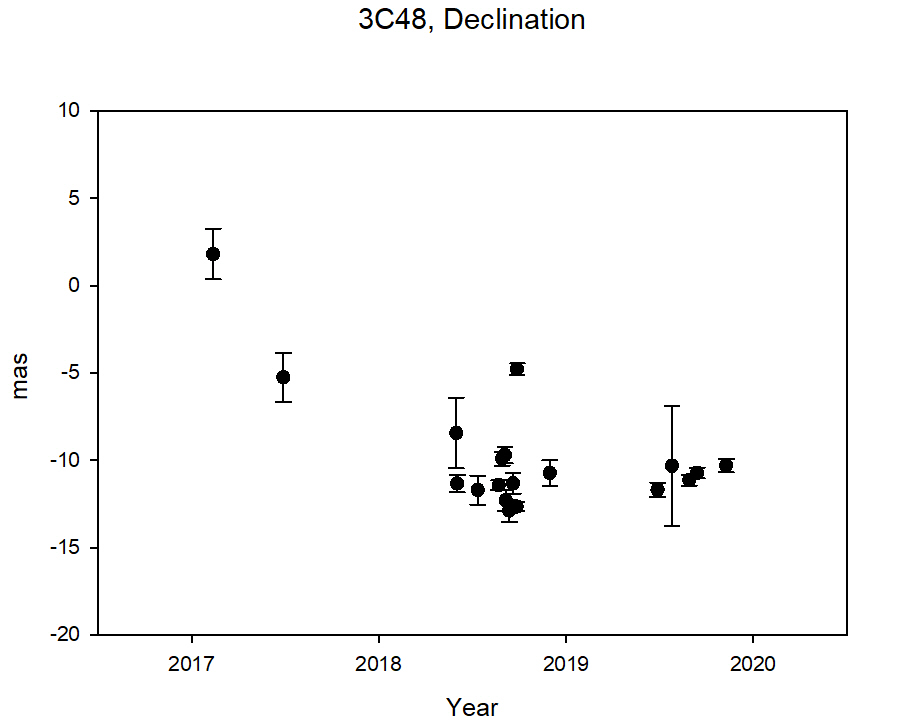}
 \caption{Relative VLBI astrometric coordinates of 3C\,48 as a function of time, zooming into the 2017–2019 section in Fig.~\ref{fig1}.} 
 \label{fig2}
\end{figure}

\subsection{3C\,48 (0134+129)}

The radio source 3C\,48 was among the first quasars identified in the optical with redshift $z = 0.367$ by \citet{1963ApJ...138...30M}. The high-resolution 8.4-GHz VLBI image of 3C\,48 obtained by \citet{2010MNRAS.402...87A} with the VLBA in 2004 displays a compact core and two bright knots in the jet, approximately 60 and 120~mas north from the core. In spite of its high luminosity at GHz frequencies, 3C\,48 is rarely observed with astrometric VLBI. The first set of 11 experiments was run in the 1990--1996, followed by an almost 20-yr break. Since 2017, 3C\,48 was monitored by IVS (Research and Development project) and VLBA networks. The monitoring of 3C\,48 with VLBA facilities is ongoing.

While the time series of right ascension does not display any remarkable change, the declination shifted by almost 55~mas between 1993 and 2018 (Fig.~\ref{fig1}). Unfortunately, due to a lack of observations, there is no chance of tracking the 3C\,48 positions during this 20-yr gap. Therefore, the positional shift of 3C\,48 towards south was noted after the ICRF3 announcement only, and this radio source is still known as astrometrically stable in the ICRF3 solution.

The two daily estimates of declination obtained during the 2017 still show a significant offset in Fig.~\ref{fig2} with respect to the estimates in 2018--2019. Therefore, these two positions could be considered as a prolongation of the long-term trend started in the 1990s (Fig.~\ref{fig1}), whereas all the new estimates in 2018 and 2019 display a barely seen astrometric variability in both components (Fig.~\ref{fig2}). That's why we have identified the epoch of ``break'' between the two data subsets near 2018.0 to estimate the positional displacement rate separately. From the sessions between 1990 and 2017, the positional displacement rate in right ascension ($\alpha$) is almost negligible,  $\mu_{\alpha} \cos\delta = 0.43 \pm 0.12$~mas\,yr$^{-1}$, but is much higher in declination ($\delta$), $\mu_{\delta} = -1.73 \pm 0.12$~mas\,yr$^{-1}$. The corresponding estimates for the period of 2018--2019 are $\mu_{\alpha} \cos\delta = 0.15 \pm 0.14$~mas\,yr$^{-1}$ and $\mu_{\delta} = -0.48 \pm 0.34$~mas\,yr$^{-1}$. This may hint that the intrinsic structure (brightness distribution) of 3C\,48 has dramatically changed until 2018, leading to the huge displacement of the VLBI reference point in the south direction. This structure has stabilized since 2018 so that the new coordinates of the source did not change over the next two years. 

Snapshot VLBA imaging data from 2018 Jul 27 (project UG002M), available in the Astrogeo database\footnote{\url{http://astrogeo.org/}, maintained by L. Petrov} and capable of revealing the most prominent compact components in the complex radio stucture of 3C\,48, indicate that the southernmost feature denoted with A1 and A2 by \citet{2010MNRAS.402...87A} has brightened significantly since 1996. Circular Gaussian brightness distribution models fitted to the calibrated visibility data in Difmap \citep{1997ASPC..125...77S} give $X$-band flux densities $S_\mathrm{A1}^{2018}=33.4$~mJy and $S_\mathrm{A2}^{2018}=144.6$~mJy for components A1 and A2, respectively. In comparison, the 1996 Jan 20 flux densities obtained by \citet{2010MNRAS.402...87A} were $S_\mathrm{A1}^{1996}=48.0$~mJy and $S_\mathrm{A2}^{1996}=18.3$~mJy. On the other hand, component B, located about 55~mas north of A1 and A2, has faded during this period, with $S_\mathrm{B}^{1996}=89.0$~mJy \citep{2010MNRAS.402...87A} and $S_\mathrm{B}^{2018}=59.6$~mJy (project UG002M).

The angular separation between the brightening southern A1+A2 components and the fading B component is in good agreement with the large displacement in declination towards the south (Table~\ref{table_radio_optical}), confirming that the astrometric displacement of 3C\,48 is directly related to the change in the source brightness distribution.
More VLBI observations of 3C\,48 are desirable to monitor further evolution of its coordinates as well as possible changes in the $\sim 1-100$~mas scale radio structure. 

\begin{figure}
\centering
 \includegraphics[width=0.4\textwidth, clip=, angle=0]{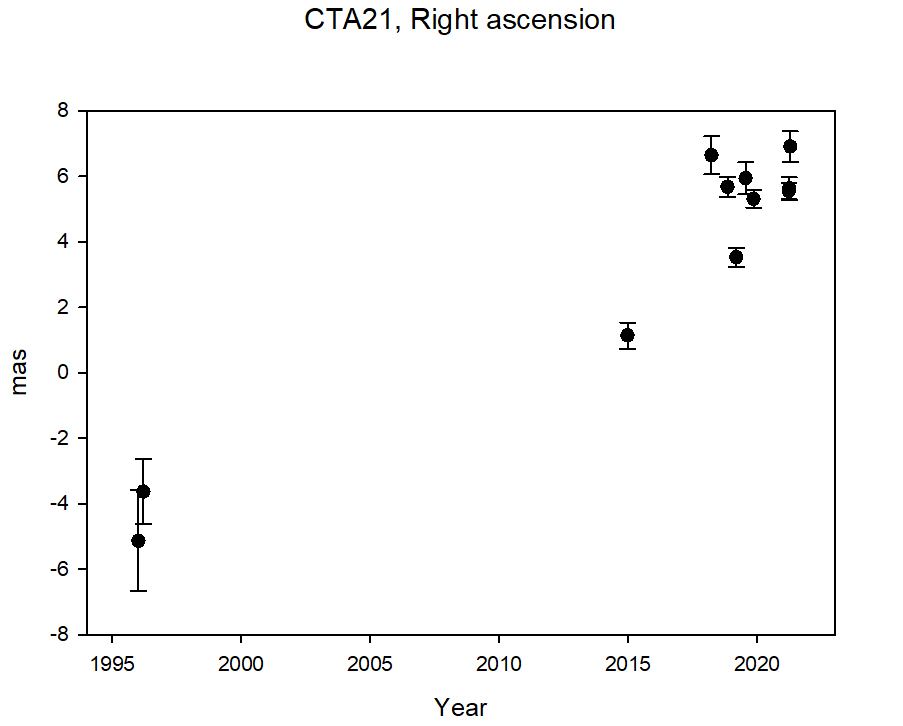}
 \includegraphics[width=0.4\textwidth, clip=, angle=0]{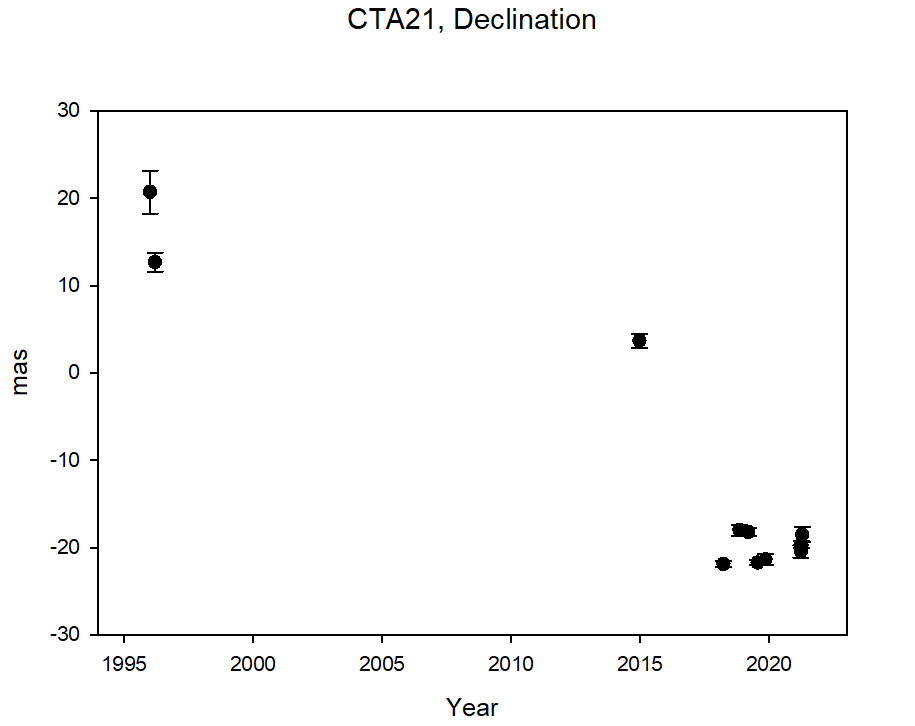}
 \caption{Time series of the relative right ascension (left) and declination (right) of CTA\,21. The (0,0) point corresponds to the solution aus2020a.crf (Table~\ref{table_source_VLBI_positions}).} 
 \label{fig3}
\end{figure}

\subsection{CTA\,21 (0316+162)}

CTA\,21 is a compact steep-spectrum and GHz-peaked spectrum source at $z = 0.907$ \citep{2007A&A...463...97L} very rarely observed in the geodetic VLBI programs. The positional offset was detected due to the two VLBA experiments in 1996 (1996 Jan 2 and 1996 Mar 13). Occasional observations by VLBA and Asia-Oceania VLBI Group for Geodesy and Astrometry (AOV) networks in the 2010s highlighted another dramatic change in its position since the first two epochs \citep{2021RNAAS...5...60F}.

A set of new VLBI experiments with CTA\,21 after the ICRF3 was released includes 2018 Mar 20 (AOV21), 2018 Nov 3 (UG002S), 2019 Mar 9 (UG003E), 2019 Jul 17 (AOV37), 2019 Nov 7 (UG003Q), 2021 Mar 20 (RUA041), and 2021 Mar 27 (RUA042). For this paper, we added results of two additional experiments in 2021 to get a new estimate of its positional displacement rate, $\mu_{\alpha} \cos\delta = 0.43 \pm 0.12$~mas\,yr$^{-1}$ in RA and $\mu_{\delta} = -1.54 \pm 0.37$~mas\,yr$^{-1}$ in Dec. (Fig.~\ref{fig3}). 

\begin{figure*}
\centering
 \includegraphics[width=0.4\textwidth, bb= 0 61 470 686, clip=, angle=0]{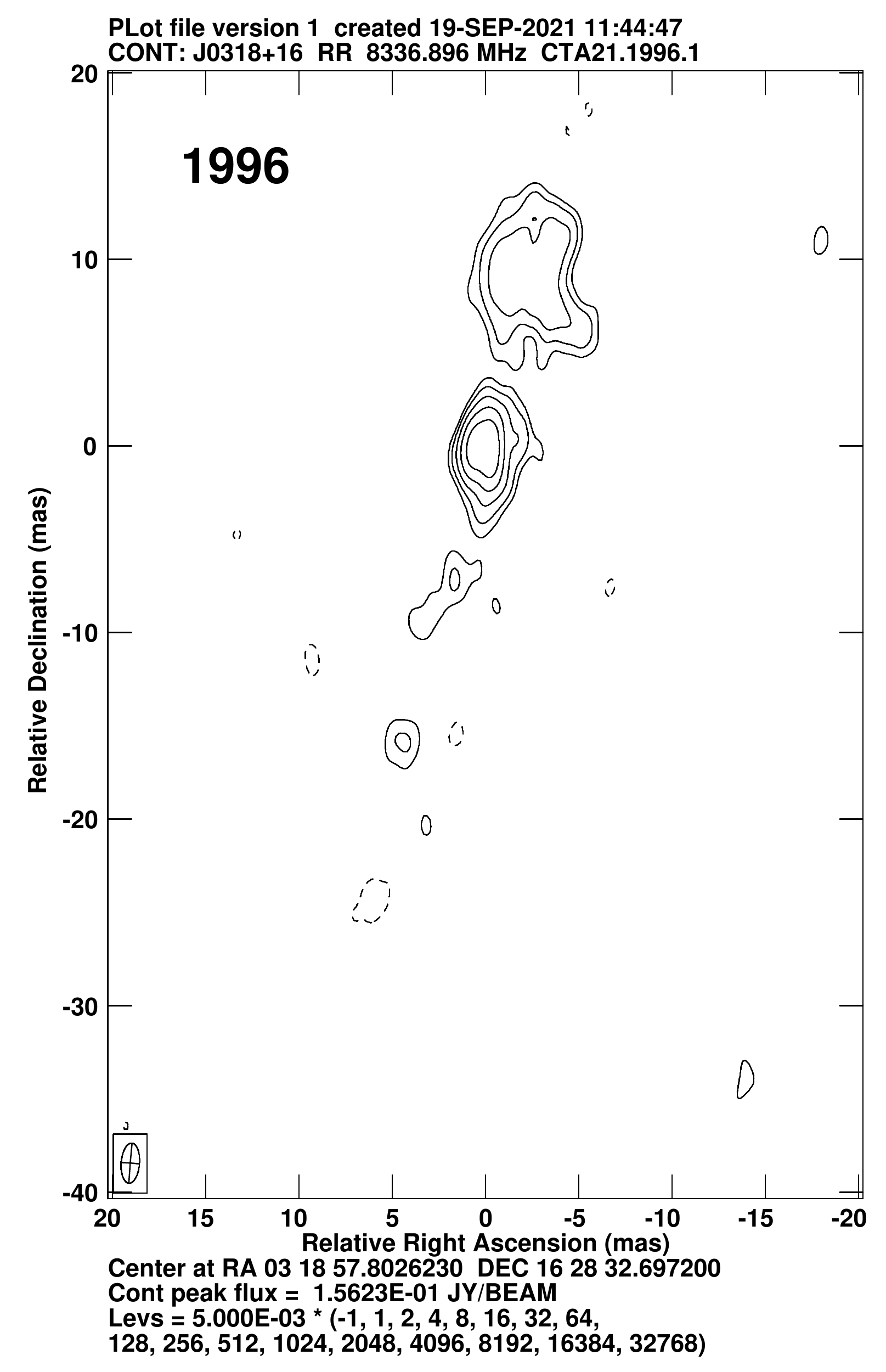}
 \vspace{3mm}
 \includegraphics[width=0.4\textwidth, bb= 0 61 470 686, clip=, angle=0]{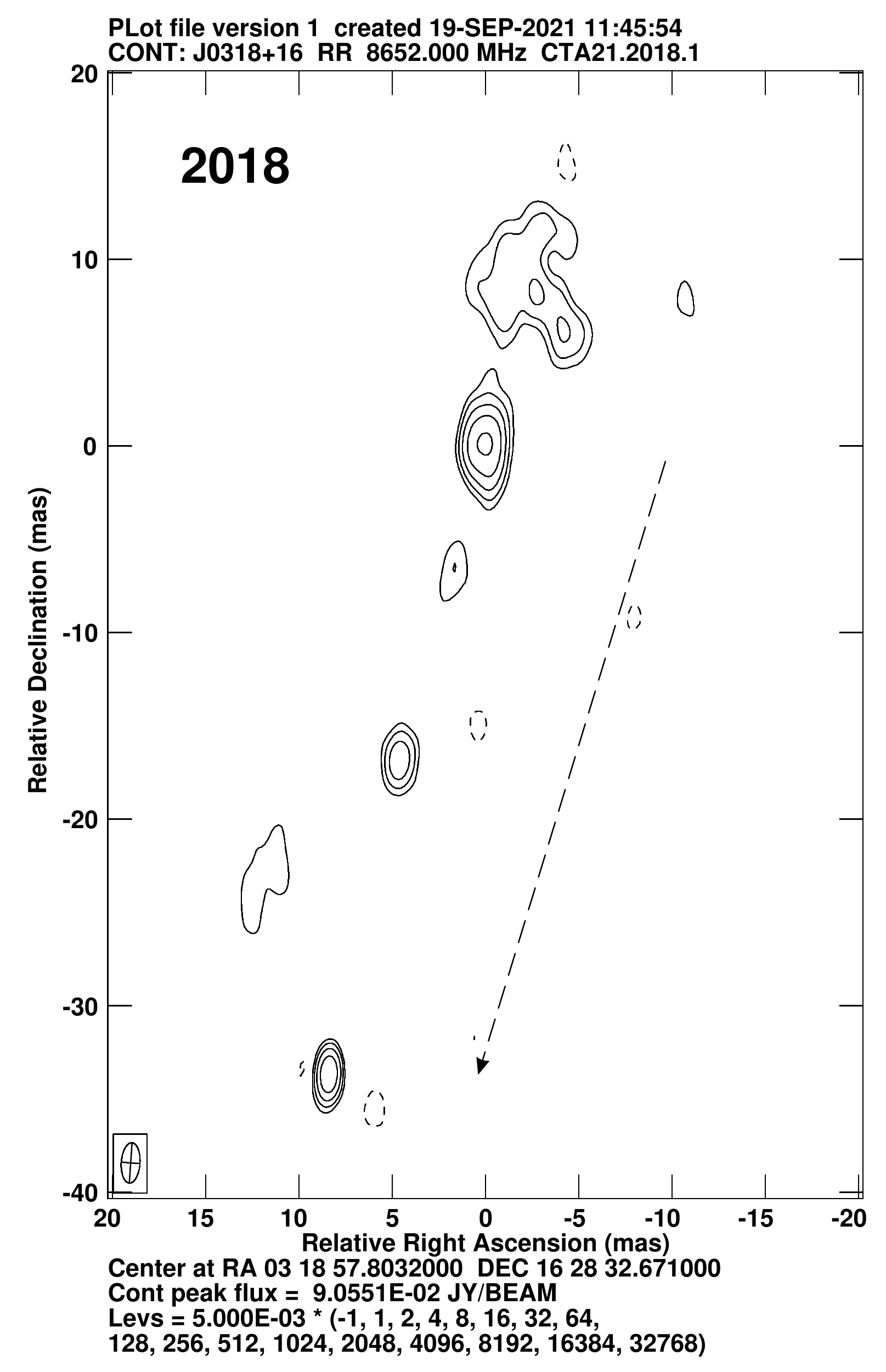}
 \caption{$X$-band VLBA images of CTA\,21 from 1996 Jan 3 and 1996 Mar 13 \citep[combined; experiment BB023,][left]{2002ApJS..141...13B} and from 2018 Nov 3 (experiment UG002S, right). The southernmost bright compact component that appears in 2018 was not detected in 1996. The arrow represents the amount and direction of the positional displacement between the two epochs as observed in the astrometric programmes (Fig.~\ref{fig3}). The lowest contours are at $\pm 5$~mJy\,beam$^{-1}$ in both images, the positive levels increase by a factor of 2. The half-power beam width is indicated in the lower-left corner. The peak brightnesses are $156$~mJy\,beam$^{-1}$ (1996) and $91$~mJy\,beam$^{-1}$ (2018). The calibrated data sets were obtained from the Astrogeo database, and the images made in Difmap \citep{1997ASPC..125...77S} and plotted in AIPS \citep{2003ASSL..285..109G}.}
 \label{cta21-img}
\end{figure*}

Despite being extensively studied, especially in the 1990s, the interpretation of the VLBI structure of CTA\,21 is still controversial \citep{2013ARep...57..423A}. \citet{2013MNRAS.433..147D} identified the AGN core with a faint component southward along the characteristic position angle of the radio structure, $\sim 125$~mas from the brightest feature of the 1.7-GHz and 5-GHz images. This component is not detected in the less sensitive snapshot images at 8.3/8.6~GHz taken in 1996 and 2018 \citep{2021RNAAS...5...60F}. On the other hand, a new unresolved bright component appeared $\sim 35$~mas south-southeast of the brightness peak, i.e. along the same position angle, by 2018. Based on its compactness and variability, it might be considered as a new core candidate. This major change in the source brightness distribution illustrated in Fig.~\ref{cta21-img} could be responsible for the significant offset detected in the astrometric position of CTA\,21. More discussion can be found in \citet{2021RNAAS...5...60F}.

\begin{figure}
\centering
 \includegraphics[width=0.42\textwidth, clip=, angle=0]{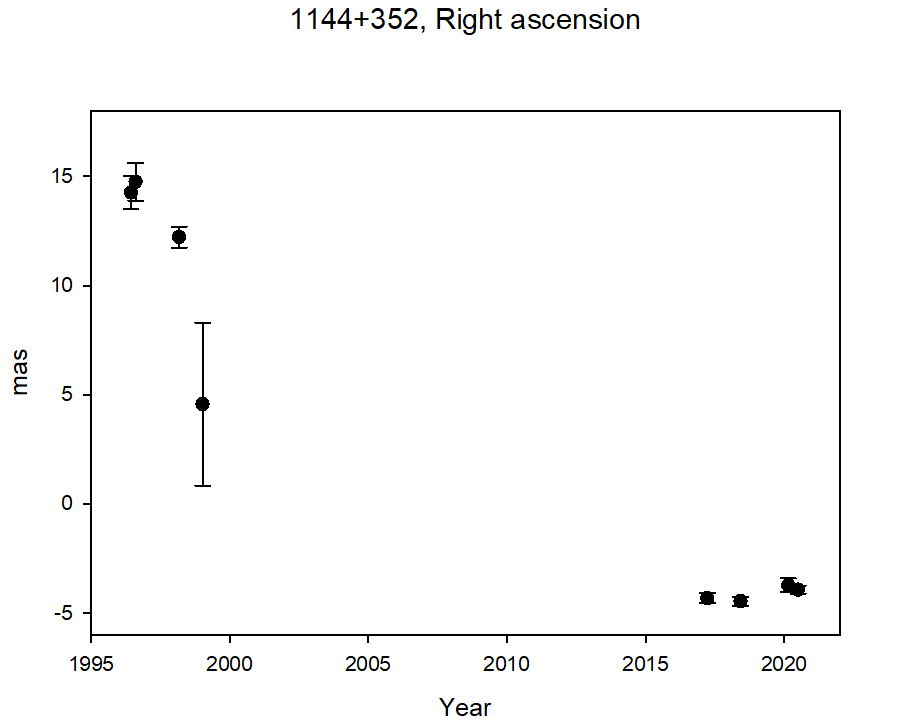}
 \includegraphics[width=0.4\textwidth, clip=, angle=0]{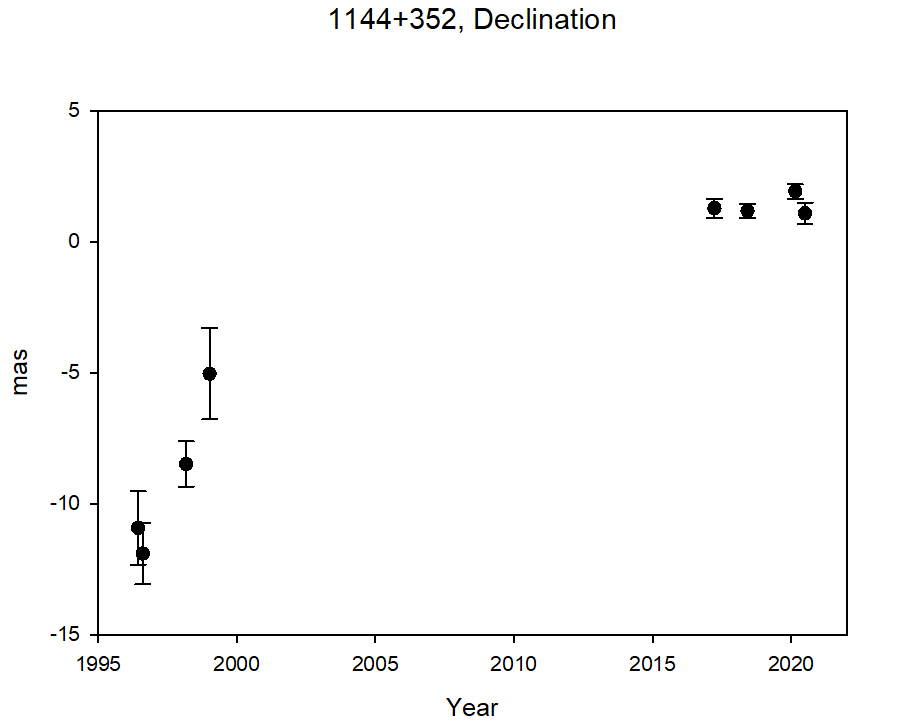}
 \caption{Relative astrometric coordinates of 1144+352 as a function of time, between 1995 and 2021. The (0,0) point corresponds to the solution aus2020a.crf (Table~\ref{table_source_VLBI_positions}).} 
 \label{fig4}
\end{figure}

\subsection{1144+352}

The radio source 1144+352 is a giant radio galaxy at low redshift ($z = 0.063$) \citep{2002MNRAS.329..700S} that was intensively studied with VLBI imaging at cm wavelengths in the past \citep[see][and references therein]{2020RNAAS...4..108T}. 
It was observed in four experiments in the 1990s (1996 Jun 7 (BB023G), 1996 Aug 7 (NAVEX S12), 1998 Mar 3 (NEOS A253), and 1999 Jan 5 (NEOS A297)) followed by an almost 20-yr gap. Only one recent experiment \citep[2017 Mar 23, UF001E,][]{2021AJ....162..121H} was included in the calculation of the position for the ICRF3, therefore, that break in the position was not detected in 2018. There are three new experiments after the ICRF3 release, two VLBA experiments, 2018 Jun 3 (UG002I) and 2020 Feb 20 (UG003U), and one RUA experiment, 2020 Jun 27 (RUA038). We only used the last three experiments to estimate the post-ICRF3 positions of 1144+352 in Table~\ref{table_source_VLBI_positions}. Since the previous publication \citep{2020RNAAS...4..108T}, we added results from two additional sessions (2020 Feb 20 VLBA and 2020 Jun 27 RUA) to confirm the offset in the position. It is not clear whether the change in the component motion was continuous or not during the gap between 1999 and 2017.

\begin{figure*}
\centering
 \includegraphics[width=0.4\textwidth, bb= 0 61 530 472, clip=, angle=0]{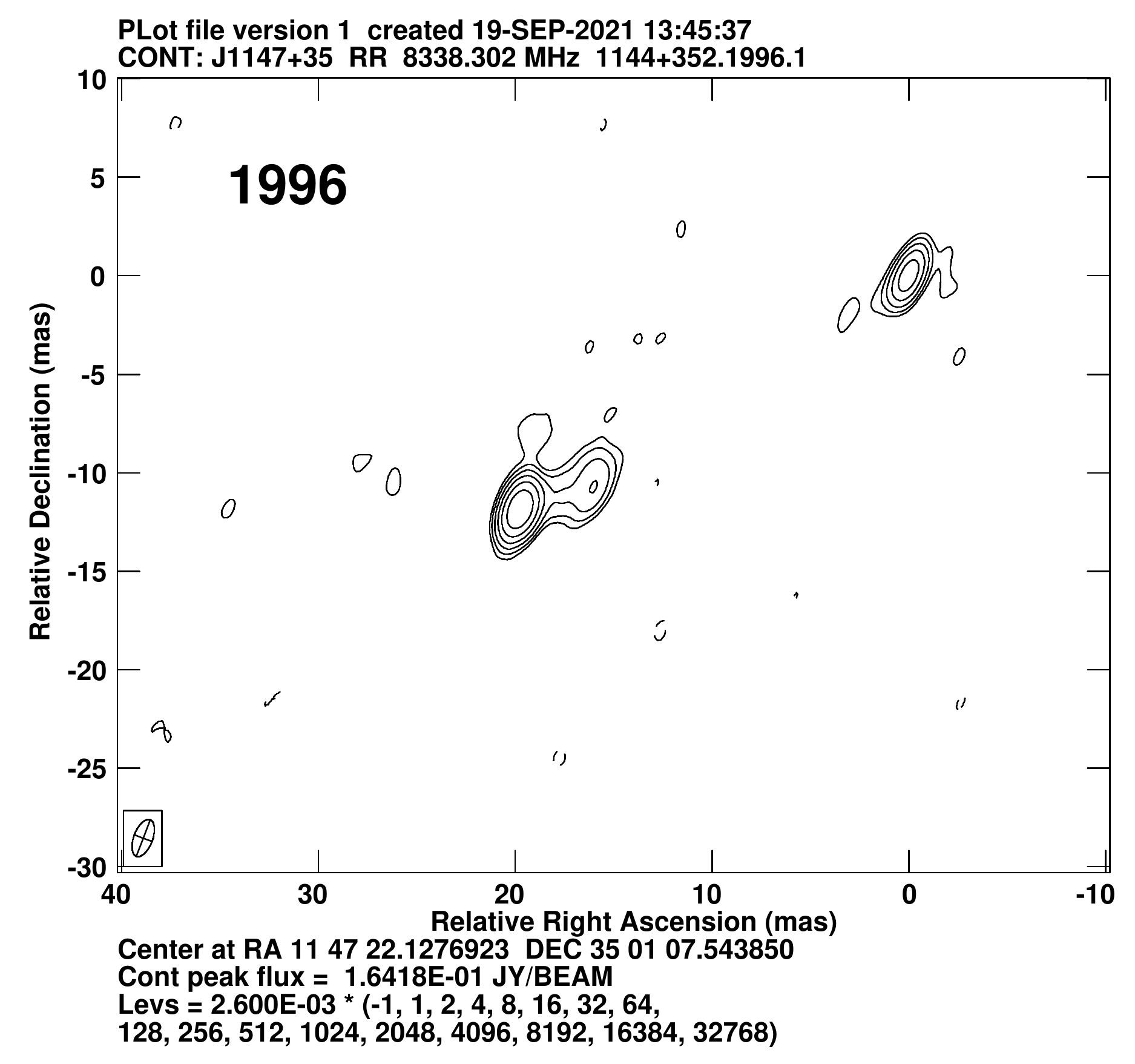}
 \vspace{5mm}
 \includegraphics[width=0.4\textwidth, bb= 0 61 530 472, clip=, angle=0]{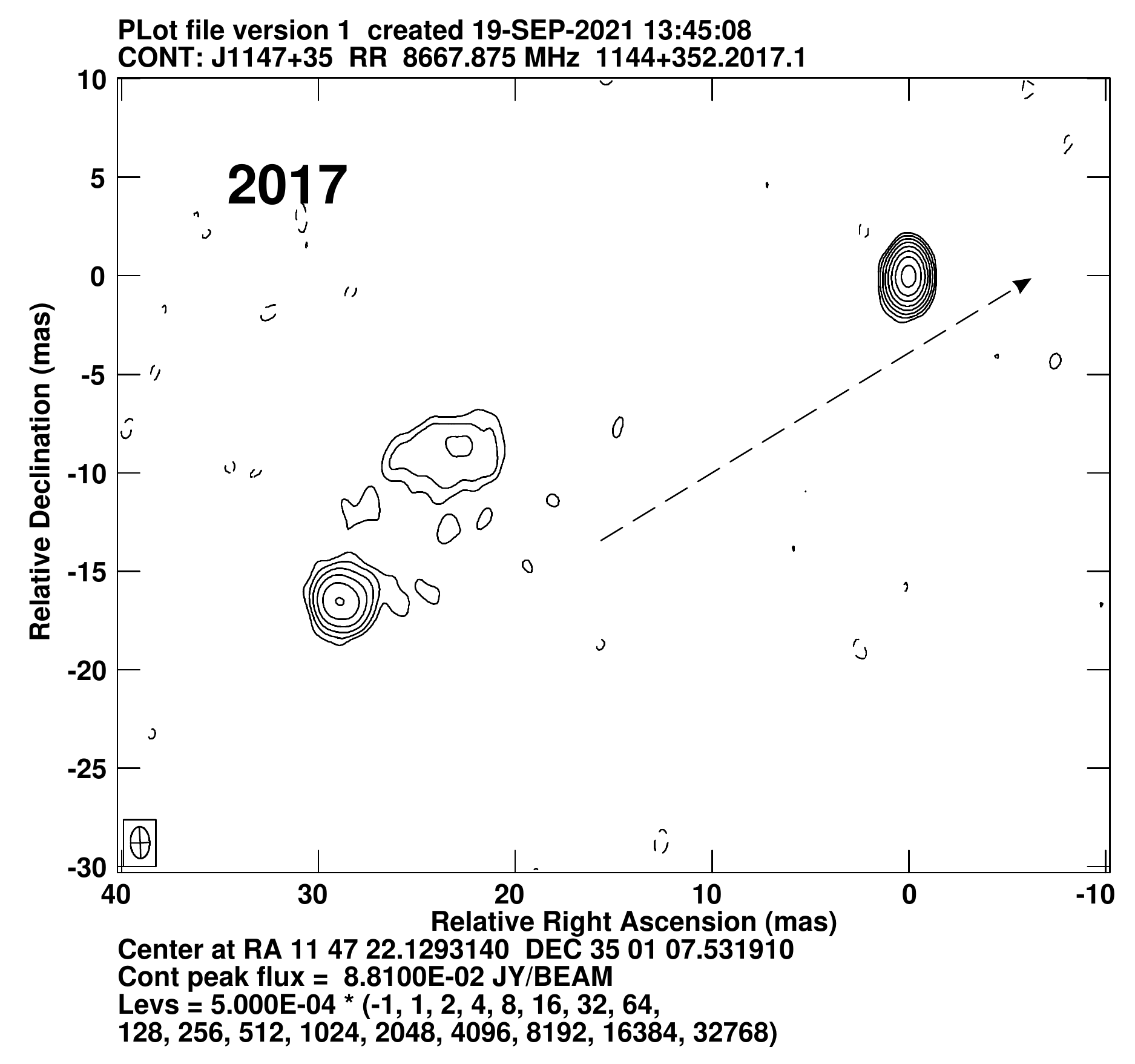}
 \caption{$X$-band VLBA images of 1144+352 from 1996 Jun 8 \citep[experiment BB023,][left]{2002ApJS..141...13B} and 2017 Mar 24 \citep[experiment UF001E,][right]{2021AJ....162..121H}. The southeastern feature was the brightest in 1996 but faded and moved away by 2017. The arrow represents the amount and direction of the positional displacement between the two epochs as observed in the astrometric programmes (Fig.~\ref{fig4}). The lowest contours are at $\pm 2.6$~mJy\,beam$^{-1}$ in the 1996 image and $\pm 0.5$~mJy\,beam$^{-1}$ in the 2017 images. The positive levels increase by a factor of 2. The half-power beam width is indicated in the lower-left corner. The peak brightnesses are $164.2$~mJy\,beam$^{-1}$ (1996) and $88.1$~mJy\,beam$^{-1}$ (2017). The calibrated data sets were obtained from the Astrogeo database, and the images made in Difmap \citep{1997ASPC..125...77S} and plotted in AIPS \citep{2003ASSL..285..109G}.} 
 \label{1144-img}
\end{figure*}

To illustrate the substantial changes in the source structure on scales of $\sim 10$~mas, we show two archival images made from data available in the Astrogeo database, one from 1996 and another from 2017 (Fig.~\ref{1144-img}). The radio core is the northwestern component \citep{1999ApJ...522..101G}. By 2017, the southestern feature, which was the brightest in 1996, became weaker than the core while they separation increased. It is reasonable to assume that, in connection with the change in the brightness distribution of 1144+352, the astrometric reference point shifted from the southestern jet component to the core. A more detailed discussion is in \citet{2020RNAAS...4..108T}.

\subsection{1328+254 (3C\,287)}

 1328+254 was classified as a compact steep-spectrum (CSS) radio source with a complex structure and several jet components \citep{1989A&A...217...44F,1990A&A...231..333F} at $z = 1.055$ \citep{1985PASP...97..932S}. \citet{1989A&A...217...44F} suggest that the core is associated to the brightest feature (component A), the jet structure is helical and shaped by precession. However, \citet{1998A&A...332...10M} and \citet{1998AJ....115.1295K} have doubts about the actual core position. \citet{1998A&A...338..840P} found that the core is not well-defined within the complex radio structure, and the jet precession model is questionable. 
 
The ICRF3 position of 1328+254 is based on two experiments before the break (one VLBA experiment, BF071B, 2002 May 14 and one joint VLBA-IVS experiment, RDV062, 2007 Mar 27) and one VLBA experiment after the break (UF001L, 2017 Jun 15).
After the ICRF3 release it was observed in two VLBA experiments, 2018 Nov 3 (UG002S) and 2019 Jun 10 (UG003L), and two RUA experiments, 2020 Nov 28 (RUA039) and 2020 Dec 26 (RUA040), therefore, we used only the data obtained from the four experiments after 2018 to estimate the post-ICRF3 positions of 1328+254 (Table~\ref{table_source_VLBI_positions}).  

This radio source yields the largest observed displacement (about 130~mas) over very short observational time (3~yr) between 2014 June and 2017 June. The enormous position shift between the 2014 and 2017 epochs is most probably linked to a brightening of a new component in the north-eastern direction from the original compact structure detected before 2014. The phase centre seems to reach its maximum offset from the 2014 position around the end of 2019 and has started an apparent motion in opposite direction in 2020 (Fig.~\ref{fig6}). By the end of 2020 it has moved on about 10 mas with respect its previous position in the mid of 2019, i.e. with the total positional displacement rate of $6-7$~mas\,yr$^{-1}$. Unfortunately, recent VLBI imaging data of 1328+254 with sufficient sensitivity are not available in the archives. New observations in the near future, before the position possibly returns to the pre-2014 value, should reveal more details about the change is the source brightness distribution likely causing the position shift.

\begin{figure}
\centering
 \includegraphics[width=0.4\textwidth, clip=, angle=0]{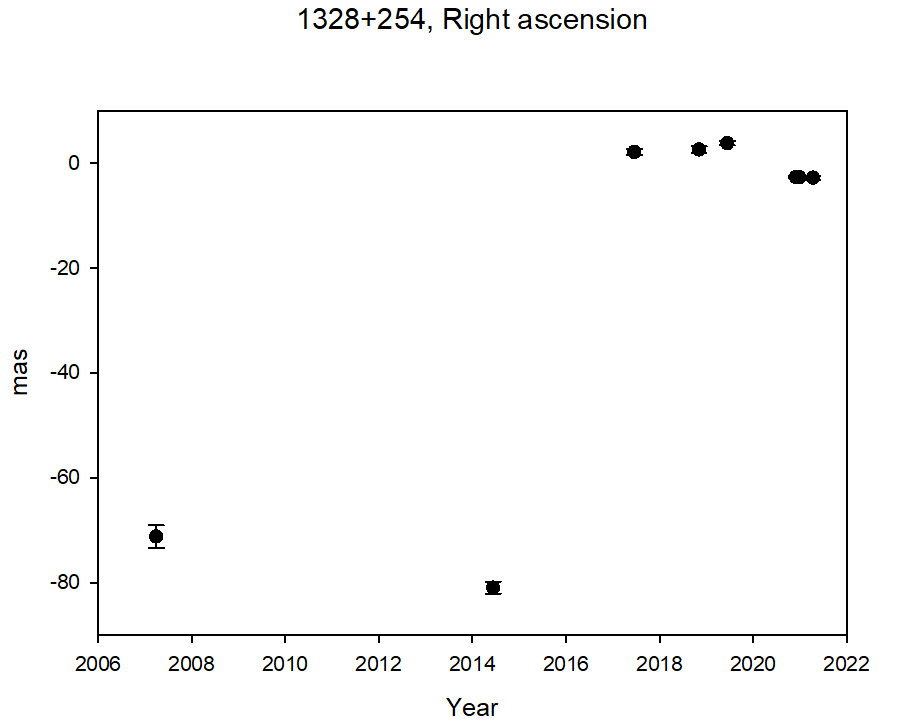}
 \includegraphics[width=0.4\textwidth, clip=, angle=0]{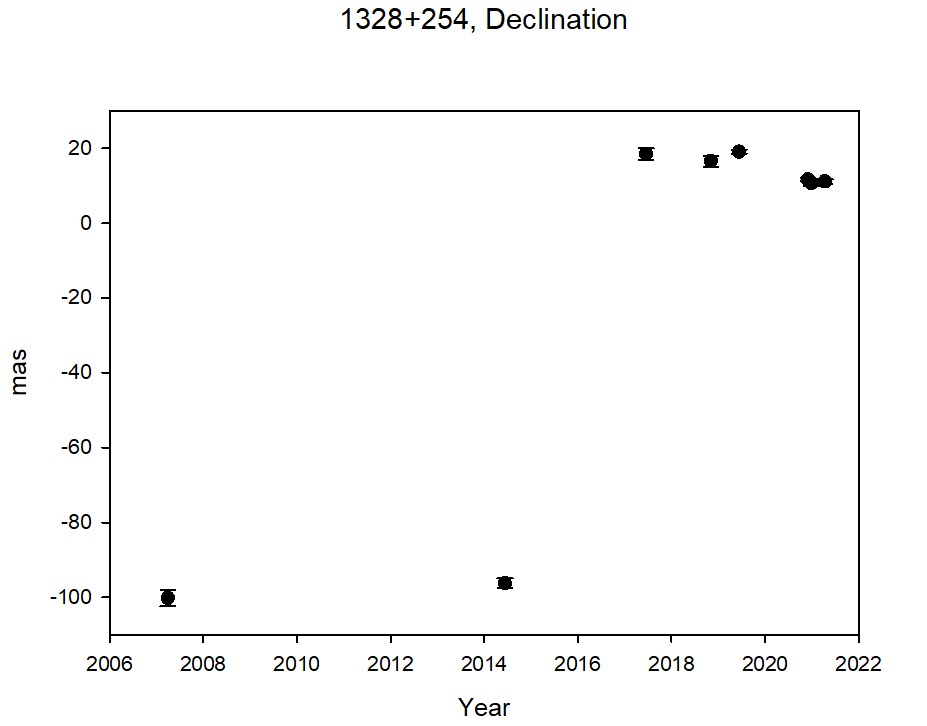}
 \caption{Relative astrometric coordinates of 1328+254 (3C\,287) as a function of time, between 2006 and 2020. The (0,0) point corresponds to the solution aus2020a.crf (Table~\ref{table_source_VLBI_positions}).} 
 \label{fig5}
\end{figure}

\begin{figure}
\centering
 \includegraphics[width=0.4\textwidth, clip=, angle=0]{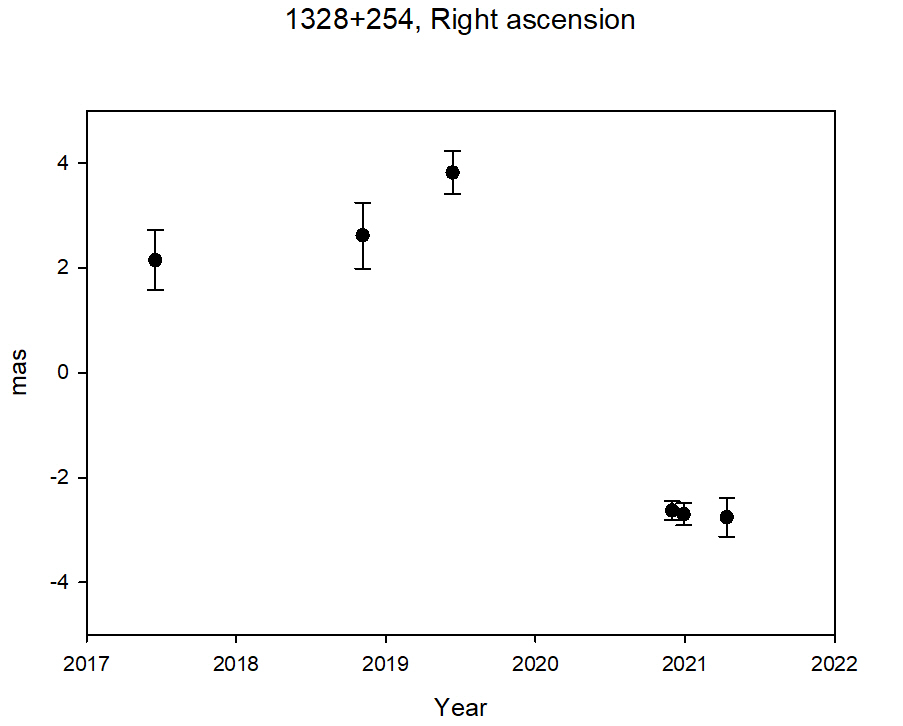}
 \includegraphics[width=0.4\textwidth, clip=, angle=0]{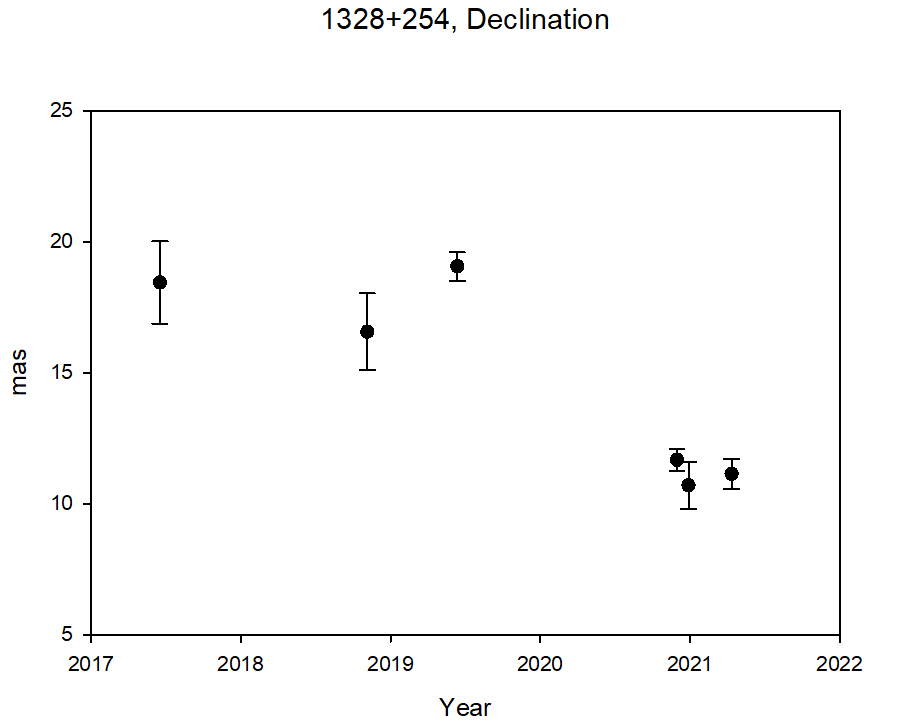}
 \caption{Relative VLBI astrometric coordinates of 1328+254 (3C\,287) as a function of time, zooming into the 2016–2021 section in Fig.~\ref{fig5}.}
 \label{fig6}
\end{figure}

\subsection{Comparison with the Paris Observatory coordinate time series}

To cross-check our results, we calculated the approximate changes in the positions of these sources based on the VLBI time series of the Paris Observatory. This solution is obtained independently with the SOLVE software \citep{1990JGR....9521991M}. Therefore, the two sets of results can be considered as free of any individual software bugs. The parameter setup in both software was done in the same way by estimating a set of the rarely observed radio source positions as daily parameters, therefore even the large corrections to the a-priori ICRF3 positions were successfully calculated. Needless to say that the parametrization of all other parameters (Earth orientation parameters, station coordinates, etc.) was realised in both software as close as possible. As the Paris Observatory solution is based on the IVS data only, and does not include the recent RUA experiments we analysed, we decided to compare the range of the positional breaks only instead of the coordinates in both solutions.  Nonetheless, the breaks in positions between the two independent solutions presented in Table~\ref{table_comparison} are consistent within 10 mas.

\begin{table}
\caption{Comparison of the positional breaks in $\Delta\alpha$ and $\Delta\delta$ calculated with two independent software, OCCAM and SOLVE}
\vspace{3mm}
\begin{tabular}{l l c c c}
 \hline
 Source & Aus2020a & & Paris Observatory & \\ 
 & RA & Dec &   RA & Dec \\ 
 &   (mas) & (mas)   &   (mas) & (mas)\\ 
  \hline \hline
  3C\,48   & -- & 45  &   --& 55   \\  [1ex] 
  CTA\,21  & 4 & 20   &   5 & 22   \\ [1ex] 
  1144+352 & 22 & 10  &  20 & 10   \\ [1ex] 
  1328+254 & 84 & 118 &  82 & 112 \\ [1ex]
 \hline
\label{table_comparison}
\end{tabular}
\end{table}

\section{Summary and conclusion}
\label{sec:sum}

We have discovered four reference radio sources that show an unprecedented change in their apparent positions  (from about 20 to 130~mas) on time scales from several years to several decades. One radio source is identified as a nearby galaxy (1144+352, $z=0.063$) while three others are common quasars at higher redshifts. This effect is predictably linked to the evolution of the intrinsic source structure and brightness distribution, although the scale of these positional changes is far beyond all astrometric instabilities known so far.

An analysis of the available VLBI images at selected observational epochs reveals that each object is composed of several compact features (a core and a long-distance jet with several knots). Simultaneous change in flux density and astrometric positions of certain components may result in such tremendous offset of the radio source phase centre which is perceived by a ground-based interferometer as an apparent motion (or positional displacement rate) of the AGNs in the sky. For all objects, these offsets are reported here for the first time, as detected from geodetic VLBI data obtained in 2018--2021, mostly with the sensitive VLBA interferometer. We should note that optical--radio positional offsets as the large as up to $\sim 1\arcsec$ were previously reported \citep[e.g][]{2013A&A...553A..13O, 2016A&A...595A...5M, Makarov2017, 2019MNRAS.482.3023P}. Those authors considered different hypotheses to explain that dramatic inconsistency, but the astrometric variability in $X$ band was not proposed among the possible interpretations. These new findings prove that such opportunity should not be ruled out.

A potential impact of such positional instability on the statistical quality of future astrometric catalogues may be severe, should the number of the radio sources with large offsets soar during the next decade. So far, only few papers considered the source structure effect on the radio source positions and all other astrometric parameters estimated with VLBI, like polar motion or the Universal Time \citep[e.g.][]{2002ivsg.conf..243S, 2018AstL...44..139T}. The major reason of the ignorance of this problem is a comparatively small contribution of the sources' structure effects ($\leq 0.1$~mas) to the total correction of the astrometric positions ($1-2$~mas). So, all the technical efforts are considered to be disproportional to the tiny improvement. Since the number of the bad behaviour radio sources in ICRF2 and ICRF3 is limited to several dozens, it appeared affordable to segregate them to a special group of the so-called specially-handled objects and not use them for substantial astrometric research and applications.

However, the finding radio sources with much stronger astrometric instability could lead to a change in this commonly used procedure. The scale of the effect is large enough to be studied even with a moderate set of observations, and the potential astrometric improvement of the reference radio source positions will secure the resources to be spent for the source structure effect correction. Assuming that the number of such radio sources grows, the attention of the IVS community to this problem will inevitably increase.

From the astrophysical point of view, there is not a universal mechanism responsible for this type of astrometric instability. Some radio sources have a typical “core--jet” structure, and the high positional displacement rate is induced by the superluminal motion of the bright jet components. Another possible scenario is a compact symmetric object (CSO) that presents two symmetric lobes (hotspots) separated on scales of up to $\sim 100$~mas beside a weak or often invisible core \citep[e.g.][]{1994ApJ...432L..87W,2012ApJ...760...77A}. The astrometric instability is likely caused by the variations of flux density in the lobes. A third mechanism is the flux density variability of the core and/or a bright compact jet component  within a complex extended radio source.

The updated positions of these four radio sources  (Table~\ref{table_source_VLBI_positions}) should be applied for the analysis of VLBI observations obtained after the break. Also, phase-referenced VLBI imaging observations and relative astrometric measurements \citep{1995ASPC...82..327B} should treat these potential calibrator sources with caution.

As all the offsets were found with recent observations made in the post-ICRF3 era, this means that any radio reference frame source may suddenly demonstrate an abrupt change in its coordinates. This seems more likely for the sources with complex, extended core--jet structure at the scale of 20--130~mas, or even more, with multiple compact components in VLBI images. Therefore, a higher number of radio sources with similar positional offsets could be found with future VLBI observations. 

We are planning to obtain new VLBI images of these targets to investigate their structure and to compare them to the existing images in the literature. Analysis of the source structure evolution is necessary to determine if it is related to the apparent displacement in the astrometric positions. 

\section*{Acknowledgements}

% Entry for the table of contents, for this guide only
%\addcontentsline{toc}{section}{Acknowledgements}

The National Radio Astronomy Observatory is a facility of the National Science Foundation operated under cooperative agreement by Associated Universities, Inc. We are also thankful for the observation time of the Warkworth 30-m radio telescope operated by the Institute for Radio Astronomy and Space Research, Auckland University of Technology. The AOV experiments were organised by the Asia-Oceania VLBI Group for Geodesy and Astrometry. The recent VLBA experiments were run by the geodetic group of the Goddard Space Flight Centre (GSFC) to monitor the radio reference frame sources with 10 VLBA antennas \citep{2021AJ....162..121H}.
The six RUA experiments were organised by the Institute of Applied Astronomy of the Russian Academy of Sciences (IAA RAS) and made use of an ad-hoc VLBI network which consists of three 32-m radio telescopes Badary, Svetloe, and Zelenchukskaya together with the 25-m radio telescope Sheshan of the Shanghai Astronomical Observatory of the Chinese Academy of Sciences (SHAO CAS), and the 40-m radio telescope of Yebes Observatory (Instituto Geogr\'{a}fico National, Spain) \citep{Frey:2019JI}.
Badary, Svetloe, and Zelenchukskaya radio telescopes are operated by the Scientific Equipment Sharing Center of the Quasar VLBI Network \citep{SHUYGINA2019150}.
%The two RUA experiments were organised by the Institute of Applied Astronomy of the Russian Academy of Sciences (IAA RAS).
%This network included the three 32-m radio telescopes of the network “Quasar” together with the 25-m radio telescope Sheshan of the Shanghai Astronomical Observatory of the Chinese Academy of Sciences (SHAO CAS), and the 40-m radio telescope of Yebes Observatory (Instituto Geogr\'{a}fico National, Spain). 
We acknowledge the use of archival calibrated VLBI data from the Astrogeo Center database maintained by Leonid Petrov. SF was supported by the Hungarian National Research, Development and Innovation Office (OTKA K134213). 

We specially thank David Gordon (US Naval Observatory) for useful discussions of the astrometric results.

This paper is published with the permission of the CEO of the Geoscience Australia.

\section*{Data Availability}

The data sets underlying this paper were derived from sources in
the public domain (in case of the IVS data \url{http://ivscc.bkg.bund.de}, in case of the Astrogeo database, \url{http://astrogeo.org}) or are available from the corresponding author upon reasonable request.

%%%%%%%%%%%%%%%%%%%%%%%%%%%%%%%%%%%%%%%%%%%%%%%%%%



\bibliography{four_quasars}
\bibliographystyle{aasjournal}

%%%%%%%%%%%%%%%%%%%%%%%%%%%%%%%%%%%%%%%%%%%%%%%%%%

% Don't change these lines
\bsp	% typesetting comment
\label{lastpage}
\end{document}